\documentclass[11pt]{article} 
\usepackage{amsmath,amsthm,amssymb,amsfonts,stmaryrd,bbold,stackrel,dsfont,mathrsfs,soul,tcolorbox,verbatimbox, tabularx, colortbl}
\usepackage{graphics}
\DeclareGraphicsExtensions{.pdf}
\graphicspath{{./}{src/}}
\usepackage{color}
\usepackage{ae,aecompl}
\usepackage[utf8]{inputenc}
\usepackage{pmboxdraw,multirow}
\usepackage[english]{babel}
\usepackage[affil-it]{authblk}
\DeclareMathAlphabet{\mathpzc}{OT1}{pzc}{m}{it}
\def\l|{\left|\left|}
\def\r|{\right|\right|}

\def\1{\mathds 1}

\definecolor{mygray}{gray}{0.8}

\usepackage[margin = 2cm]{geometry}

\title{Burns turbulent dispersion \\ considers the dispersed phase as a passive scalar}
\author{Corentin Reiss}
\affil{Universit\'e Paris-Saclay, CEA, Service de Thermo-hydraulique et de M\'ecanique des Fluides, \\Gif-sur-Yvette, 91191, France}
\affil{Institut de Mécanique des Fluides de Toulouse, Université de Toulouse, CNRS, INPT, UPS,\\
	All\'ee du Prof. Camille Soula, Toulouse, 31400, France}

\date{February 2024}

\begin{document}
\maketitle

\begin{abstract}
The Burns turbulent dispersion force is the most commonly used turbulent dispersion in the two-fluid RANS bubbly-flow literature.
However, its derivation is based on a series of hypothesis that are difficult to justify in industrial flows.
It is shown that in low-void fraction vertical pipe flow, the Burns turbulent dispersion formulation is equivalent to considering the radial movement of the gas phase like a passive scalar that follows turbulent eddies.
This is not apparent in the derivation of the force.
As bubbles in pressurized water reactor (PWR) conditions have a low Stokes number ($\sim 10^{-1}$), considering bubbles are transported by turbulence is a good approximation for bubble dispersion in pipe flow.
Therefore, the Burns turbulent dispersion force is appropriate to represent bubble dispersion in low-void fraction PWR flows.
\end{abstract}

\section{Introduction}

Notation conventions:
\begin{itemize}
	\item All forces written here are applied to the gas phase
	\item We define $X'$ the fluctuations of a quantity $X$, and write $u_i = u_i + u_i'$, $P_i = P_i + P_i'$, $\alpha_i = \alpha_i + \alpha_i'$... 
\end{itemize}

\paragraph{General form of turbulent dispersion}

To correctly predict the flow pattern in nuclear applications, one must determine the impact of \textbf{liquid turbulence} on bubble movement.

In 6-equation component and system-scale codes, a channel or subchannel turbulent dispersion can be determined from the void fractions and relative phase velocities~\cite{Emonot2011}. 

In 4-equation component-scale codes, a void diffusion coefficient or a void drift velocity can be calibrated on experimental data, as in CTF~\cite{CTF2023}. 
These expressions depend on the flow velocity, which is closely related to the turbulent intensity of the flow.

For 6-equation CFD codes, the formulations found in the literature mostly derive from theoretical analysis of bubble-fluid interactions. 
The turbulent dispersion generally reads~\cite{Burns2004}:
\begin{equation}
\vec{F}_{TD}=- C_{TD} \rho_l k \underline{\nabla} \alpha_g
\end{equation}
Where $C_{TD}$, the turbulent dispersion coefficient, is a dimensionless coefficient that depends on local flow conditions, $\rho_l$ the liquid density, $k$ the turbulent kinetic energy and $\alpha_g$ the gas void fraction.

\paragraph{RANS expression of the mass equation}

The instantaneous mass equation for the gas phase in a two-fluid adiabatic model reads~\cite{HibikiIshii2006}:
\begin{equation}
	\frac{\partial  \chi \rho_g}{\partial t} + \nabla \cdot ({\chi \rho_g} \vec{u_g}) = 0
\end{equation}
With $\chi$ the phase indicator function, $\rho_g$ the gas density and $\vec{u_g}$ the gas velocity.

Time-averaging this equation (and not phase-averaging it) and separating the contributions leads to:

\begin{equation}
	\frac{\partial  <\chi \rho_g>}{\partial t} + \nabla \cdot ( <\chi \rho_g> <\vec{u_g}> + <\chi \rho_g \vec{u_g}'>) = 0
\end{equation}

If one uses the eddy-viscosity hypothesis applied to void fraction~\cite{Burns2004}: $<\chi \rho_g \vec{u_g}'> = - \rho_g \nu_\alpha \nabla \alpha_g$, with $<\chi \rho_g> = \alpha_g \rho_g$, and $\nu_\alpha$ a diffusion coefficient for the void fraction, we obtain a convection-diffusion equation on the void fraction:

\begin{equation}
	\frac{\partial  \alpha_g \rho_g}{\partial t} + \nabla \cdot ( \alpha_g \rho_g <\vec{u_g}>)  = \nabla \cdot (\rho_g \nu_\alpha \nabla \alpha_g)
\end{equation}

\textbf{Favre averaging} is different from time averaging. 
The Favre average, also called statistical average and phase average, of a quantity $\Phi$ is a mass and void fraction weighted average: $<\Phi>_{Favre}=\frac{<\Phi\alpha\rho>_{Time}}{<\alpha\rho>_{Time}}$. 
Favre-averaging the mass equation results in a convection equation on void fraction:
\begin{equation}
	\frac{\partial  \alpha_g \rho_g}{\partial t} + \nabla \cdot ( \alpha_g \rho_g <\vec{u_g}>_{Favre})  = 0
\end{equation}

The Favre average is usually used when going towards a RANS formulation~\cite{HibikiIshii2006}.

\paragraph{Importance of the turbulent time scale and dissipation rate}

The turbulent dispersion of a passive scalar is a stochastic process. 
We place ourselves in a $k-\omega$ turbulence framework~\cite{Kok1999}, with $\omega$ the turbulent dissipation rate.
A dye particle will follow eddies that move at a velocity $\sim \sqrt{k}$ for a time $\tau_{turb}=\frac{1}{\omega}=\frac{C_\mu k}{\epsilon}$~\cite{Ktau2000}.
The diffusion is characterized by a turbulent viscosity $\nu_t = k\tau_{turb}$.
It is therefore surprising that most classical formulations for bubble dispersion hide $\tau_{turb}$ in the turbulent dispersion coefficient even though it is sure to play a part.

The DEBORA experiment was built at CEA to simulate pressurized water reactor (PWR) conditions~\cite{Debora2001}. 
In CFD simulation of this experiment, $\omega$ is typically $10^{2}-10^{3}\text{s}^{-1}$. The turbulent time scale is therefore $\tau_\text{turb} \sim 10^{-3}-10^{-2}\text{s}$.

Furthermore, the bubble response time will be in $\alpha_g\rho_g u_g/F_\text{drag}$, where $F_\text{drag}$ is the drag force.
So $\tau_\text{bubble} \sim C_D d_b \rho_g/(\rho_l u_r)$ with $C_D$ the drag coefficient and $d_b$ the bubble diameter~\cite{Tomiyama1998}. 
In nuclear reactor conditions, $\rho_g/\rho_l\sim0.1$, $d_b\sim 5\cdot 10^{-4}$m, $u_r\sim 0.1m/s$, $C_D\sim 1$. 
Therefore, $\tau_\text{bubble} \sim 5 \cdot 10^{-5} $, $St_\text{bubble} \sim 10^{-2} $.

However, one must take into account the virtual mass force on the bubbles.
The bubble inertia then increases and $\tau_\text{bubble} \sim C_{VM} C_d d_b /u_r$. 
If we take $C_{VM}\sim .5$~\cite{Zuber1964},  $\tau_\text{bubble} \sim  2 \cdot 10^{-4}$, which reduces the difference with the turbulent time scale.  
$St_\text{bubble} \sim 10^{-1} $, still smaller than 1.
Bubbles should then follow the liquid flow in the radial direction.
In the axial direction, buoyancy creates a velocity difference.

\section{Formulations from the literature derived from averaging}

This sections covers the formulations of Burns~\cite{Burns2004} and Laviéville~\cite{EDFGTD2017}. 
These methodologies are used indiscriminately to describe bubble and particle dispersion.

\paragraph{Burns et al.}

The authors start from the two-phase momentum equation:
\begin{equation}
	\frac{\partial \alpha_k \rho_k \vec{u_k}}{\partial t} 
	+ {\nabla} \cdot ({\alpha_k \rho_k \vec{u_k}} \otimes \vec{u_k})
	= 
	-\alpha_k {\nabla} P +
	{\nabla} \cdot [ \alpha_k \mu_k {{\nabla}} \vec{u_k}]
	+ \vec{F}_{ki}
	+ \alpha_k\rho_k\vec{g}
\end{equation}
Where $k$ can be either the liquid or the gas phase, $\vec{F}_{ki}$ are the interfacial forces, $\vec{g}$ the gravitational acceleration and $\mu_k$ the dynamic viscosity.

They then apply a Reynolds-Averaged Navier-Stokes methodology to the two-phase flow. 
They consider that the unstable contribution to the drag force is the only one that is non-negligible after averaging. 
Burns et al. write:
\begin{equation}
	\vec{F}_\text{TD} =
	<\frac{3}{4}C_D\frac{\alpha_g\rho_l}{d_S}|\vec{u_g}-\vec{u_l}|>
	\left(
	\frac{<\alpha_g'u_g'>}{\alpha_g}
	- \frac{<\alpha_l'u_l'>}{\alpha_l}
	- \frac{<a_i'(u_g'-u_l')>}{<a_i'>}
	\right)
\end{equation}
Where $d_S$ is the bubble Sauter mean diameter.

They then apply an eddy viscosity hypothesis ($<\alpha_k'u_k'>=-\frac{\nu_t}{Pr_t}\nabla \alpha_k$ with $\nu_t$ the turbulent viscosity and $Pr_t$ the bubble turbulent Prandlt number) to arrive at their final formulation for two-phase flow:
\begin{equation}
	\label{eq_burns_force}
	\vec{F}_\text{TD} =
	- <\frac{3}{4}C_D\frac{\alpha_g\rho_l}{d_S}|\vec{u_g}-\vec{u_l}|>
	\frac{\nu_t}{Pr_t}
	\left(
	\frac{1}{\alpha_g}
	+ \frac{1}{\alpha_l}
	\right)
	\nabla \alpha_g
\end{equation}

However, this formulation isn't usable as such as there is still an averaged value inside. 
The authors then do the unjustified following hypothesis to close the turbulent dispersion force:
\begin{equation}
	 \label{eq_approx_avg}
	 <\frac{3}{4}C_D\frac{\alpha_g\rho_l}{d_S}|\vec{u_g}-\vec{u_l}|>
	 =  \frac{3}{4}C_D\frac{\alpha_g\rho_l}{d_S}|\vec{u_g}-\vec{u_l}|
\end{equation}
Where the $C_D$ used is that af the drag force of steady-state bubbles rising in the liquid $|\vec{u_g}-\vec{u_l}|$. 
\\~\\
For this formulation to be valid, the following hypothesis must be met:
\begin{itemize}
	\item The only non-negligible unstable contribution to the force balance is through the drag force. 
	Other sources disagree with this hypothesis~\cite{EDFGTD2017, Chahed2019}.
	\item The drag force formulation used in the fluctuating flow before the RANS step must be valid. 
		A steady-state single-bubble formulation is most often used. 
		This means in particular that:
	\begin{itemize}
		\item The drag coefficient must be the same in the vertical and other directions, which is not necessarily the case for deformed bubbles
		\item The typical size of turbulent eddies is much larger than the size of bubbles, so they are faced with a uniform velocity field
		\item The typical time scale of eddies dispersing the bubbles is much larger than the time required for the drag force from the eddy on the bubble to reach its stationary value
	\end{itemize}
	\item The approximation used in equation~\ref{eq_approx_avg} is valid
\end{itemize}

These hypothesis are not necessarily met in high-void fraction fast-flowing conditions.

\paragraph{Laviéville et al.}

In this work, the turbulent dispersion formulation comes from an averaging of the lagrangian equation of motion for a single bubble which is then transformed back in eulerian coordinates.

The authors take into account many different forces acting on the bubbles: added mass, drag, lift and pressure variations (which, as with the application of the Burns methodology, must have the same formulation in a fluctuating flow as in a stationary one).

Without going into detail as to how their formulation is derived, they make the following hypothesis:
\begin{itemize}
	\item The fluctuations of single bubbles in a large amount of fluid are identical to those of a bubble swarm
	\item The formulations on lift, drag and added mass forces evaluated on single bubbles in an undisturbed bulk used before RANS averaging are valid in unsteady high-void fraction flows
	\item We can build a time scale of fluid turbulence along bubble trajectories and one for drag
	\item $<u_g'u_g'>$ and $<u_g'u_l'>$ can be related to $<u_l'u_l'>$ in all flow conditions through complex formulas that use the aforementioned time scales and an added mass coefficient
\end{itemize}

As above, these hypothesis are not necessarily met in high-void fraction fast-flowing conditions.

\section{Burns formulation considers bubbles as a passive scalar}

\paragraph{Force equilibrium in a bubbly boiling pipe flow}

In the core region of bubbly boiling PWR-condition pipe flows, the radial lift and added mass forces, bubble inertia and the pressure gradient play a minor role~\cite{Reiss2023}.
The radial bubble movement is piloted by the balance between the turbulent dispersion and the drag force. 
This can be seen by post-processing forces from CFD simulations.
Furthermore, the axial velocity is much larger than the radial velocity. The radial drag force reads:
\begin{equation}
	\vec{F}_{D,r} = - \frac{3}{4}C_D\frac{\alpha_g\rho_l}{d_S}|\vec{u_{g,z}}-\vec{u_{l,z}}|({u_{g,r}}-{u_{l,r}})
\end{equation} 

Using the Burns formulation from equation~\ref{eq_burns_force}, we obtain: \begin{equation}
	\frac{3}{4}C_D\frac{\alpha_g\rho_l}{d_S}|\vec{u_{g,z}}-\vec{u_{l,z}}|({u_{g,r}}-{u_{l,r}}) 
	+ \frac{3}{4}C_D\frac{\alpha_g\rho_l}{d_S}|\vec{u_g}-\vec{u_l}|
	\frac{\nu_t}{Pr_t}
	\left(
	\frac{1}{\alpha_g}
	+ \frac{1}{\alpha_l}
	\right)
	\partial_r \alpha_g
	= 0
\end{equation} 
Therefore, the relative velocity of the vapor phase is:
\begin{equation}
	{u_{g,r}}-{u_{l,r}} = -
	\frac{\nu_t}{Pr_t}
	\left(
	\frac{1}{\alpha_g}
	+ \frac{1}{\alpha_l}
	\right)
	\partial_r \alpha_g
	\label{eq_ur}
\end{equation} 

\paragraph{Simplified situation}

We will now place ourselves in the following situation:
\begin{itemize}
	\item The bubbles are in homogeneous isotropic liquid turbulence, with a turbulent diffusivity $\nu_t$
	\item The bubble movement is piloted by the drag-dispersion equilibrium
	\item There is no phase change or bubble injection in the area of interest
	\item The flow is stationary
	\item The volume masses $\rho_l$ and $\rho_g$ are constant
	\item The vertical velocities are uniform. 
\end{itemize} 
This seemingly far-fetched situation is that of a vertical bulk liquid flow in a turbulent large tank where bubbles are injected upstream of the area of interest, as in the Alméras et al. experiments~\cite{Almeras2019}.
It is also not too different from the core of a saturated bubbly flow.
\\~\\
As the flow is stationary and has no phase change:
\begin{equation}
	\label{eq_div_0}
	\nabla \cdot (\alpha_k \vec{u_k}) = 0
\end{equation}

We will now place ourselves in a 2D channel, i.e. a pipe with a rectangular cross-section~\cite{Martin1972}.
We name $z$ the vertical axis and $x$ the horizontal axis along the small width. 
This leads to us solving the following system of equations:
\begin{equation}
	\label{eq_system}
	\begin{cases}
		(1) & u_{gz}\partial_z\alpha_g  + \partial_x(\alpha_g u_{gx})=0 \\
		(2) & -u_{lz}\partial_z\alpha_g + \partial_x((1-\alpha_g) u_{lx})=0 \\
		(3) & {u_{g,x}}-{u_{l,x}} = - \frac{\nu_t}{Pr_t} \left(\frac{1}{\alpha_g}	+ \frac{1}{1-\alpha_g} \right) \partial_x \alpha_g
	\end{cases}       
\end{equation}

\paragraph{Integration}

By replacing $u_{lx}$ by $u_{gx} - (u_{gx}-u_{lx})$ in $(2)$ and integrating equations~$(1)$ and $(2)$ above between $0$ and $X$ we obtain:
\begin{equation}
	\begin{cases}
		(1) & u_{gz}\int_{0}^{X}(\partial_z\alpha_g)dx  + \alpha_g(X) u_{gx}(X)    - \alpha_g(0) u_{gx}(0) =0  \\
		(2) & -u_{lz}\int_{0}^{X}(\partial_z\alpha_g)dx + (1-\alpha_g(X)) u_{gx}(X)- (1-\alpha_g(0)) u_{gx}(0) \\
		~ & =(1-\alpha_g(X))(u_{gx}(X)-u_{lx}(X)) - (1-\alpha_g(0))(u_{gx}(0)-u_{lx}(0))
	\end{cases}       
\end{equation}

We then do $(1)\cdot(1-\alpha_g) - \alpha_g \cdot (2)$, inject $(3)$ and do another derivation. We obtain:
\begin{equation}
	\left( (1-\alpha_g) u_{gz} + \alpha_g u_{lz}\right) \partial_z \alpha_g 
	= \partial_x\left(\frac{\nu_t}{Pr_t}\partial_x \alpha_g\right)
	+ \partial_x \alpha_g \left( u_{gz}-u_{lz}\right)\int_{0}^{X} \partial_z \alpha_g dx
	\label{eq_with_integral}
\end{equation}
The left-hand term and first right-hand term give us a radial diffusion equation in the convected fluid. 

\paragraph{Big approximation}

We suppose that axial velocity differences have little impact on the transverse flow.
In other words, $u_{gz}, u_{lz} \gg u_{gz}-u_{lz}$.
As $u_{gz}-u_{lz}\sim0.1$m/s in bubbly flow~\cite{Sugrue2017}, this approximation is valid in a fast-flowing rectangular channel ($u_{lz}>1$m/s).
This yields $\alpha_g u_{gx}\sim-(1-\alpha_g) u_{lx}$ from equation \ref{eq_system}-$(1)$ and \ref{eq_system}-$(2)$. 
Equation \ref{eq_system}-$(3)$ then becomes:
\begin{equation}
 {u_{g,x}}-{u_{l,x}} = {u_{g,x}}+{u_{g,x}}\frac{\alpha_g}{1-\alpha_g} =\frac{u_{g,x}}{1-\alpha_g}
 = - \frac{\nu_t}{Pr_t} \left(\frac{1}{\alpha_g}	+ \frac{1}{1-\alpha_g} \right) \partial_x \alpha_g
\end{equation}
So:
\begin{equation}
	\alpha_g u_{g,x} = - \frac{\nu_t}{Pr_t}\partial_x \alpha_g
\end{equation}
Using equation \ref{eq_system}-$(1)$, we obtain:
\begin{equation}
	\label{eq_diffusion}
	\left( (1-\alpha_g) u_{gz} + \alpha_g u_{lz}\right) \partial_z \alpha_g 
	= \partial_x\left(\frac{\nu_t}{Pr_t}\partial_x \alpha_g\right)
\end{equation}

This is the same equation as above, without the integral term: a radial diffusion equation in the convected fluid, with a diffusivity equal to the liquid turbulent viscosity. 

This means that the integral term in equation~\ref{eq_with_integral} represents the contribution of axial velocity difference between the two phases to the evolution of $\alpha_g$.
This is coherent with the fact that this term contains $u_{gz}-u_{lz}$ and $\partial_z \alpha_g$.

\paragraph{Big approximation in cylindrical coordinates}

We now try to expand our previous reasoning in cylindrical coordinates. The system we now must solve is:
\begin{equation}
\label{eq_system_cylinder}
\begin{cases}
	(1) & u_{gz}\partial_z\alpha_g  + 1/r\partial_r(r \alpha_g u_{gr})=0 \\
	(2) & -u_{lz}\partial_z\alpha_g +  1/r\partial_r(r (1-\alpha_g) u_{lr})=0 \\
	(3) & {u_{g,r}}-{u_{l,r}} = - \frac{\nu_t}{Pr_t} \left(\frac{1}{\alpha_g}	+ \frac{1}{1-\alpha_g} \right) \partial_r \alpha_g
\end{cases}       
\end{equation}

We also place ourselves in fast-flowing pipe flow, where  $\alpha_g u_{gr}\sim-(1-\alpha_g) u_{lr}$. As above, $\alpha_g u_{g,x} = - \frac{\nu_t}{Pr_t}\partial_x \alpha_g$.
And:
\begin{equation}
	\label{eq_diffusion_cylinder}
	\left( (1-\alpha_g) u_{gz} + \alpha_g u_{lz}\right) \partial_z \alpha_g 
	= 1/r\partial_r\left(r \frac{\nu_t}{Pr_t}\partial_r \alpha_g\right)
\end{equation}

Which is also a diffusion equation.

\paragraph{Simulation}

We test the hypothesis that the Burns turbulent dispersion force disperses bubbles as if they were passive scalars in industrial geometries.

We simulate the following situation: a tube where bubbles are injected at the core~\cite{Colin2002}.
The multi-phase module of the TrioCFD code is used~\cite{Angeli2015}.
The tube is 4cm in diameter, 4m long and a small amount of air is injected 2m downstream of the entrance.
Bubbles are $0.92\text{mm}$ in diameter, and $Re_b\sim 100$.
The liquid bulk velocity is 1m/s and a $k-\omega$ turbulence model is used.
The void fraction predicted by the code along the width of the pipe at different positions downstream of the inlet is compared.

The first simulated configuration is a 2-fluid model with drag, added mass, gravity and Burns turbulent dispersion forces. 
The lift force and a wall correction were left out to compare the Burns force with simple diffusion

The second simulated configuration is a mixture model where the relative velocity $\vec{u_g}-\vec{u_l}$ can be enforced in the code~\cite{Bois2021mixture_model}. 
Here, we use:
\begin{equation}
	\vec{u_g}-\vec{u_l} = u_{r,\text{Ishii-Zuber}} - \frac{\nu_t}{Pr_t \alpha_g} \nabla \alpha_g
\end{equation}
This amounts to directly enforce equation~\ref{eq_ur}, and obtain an effective diffusion equation on the vapor velocity.
The momentum and mass equations for the mixture are then solved by the code.

\begin{figure}[h]
	\begin{center}
	\label{fig_burns_diffu}
	\includegraphics[width=8cm]{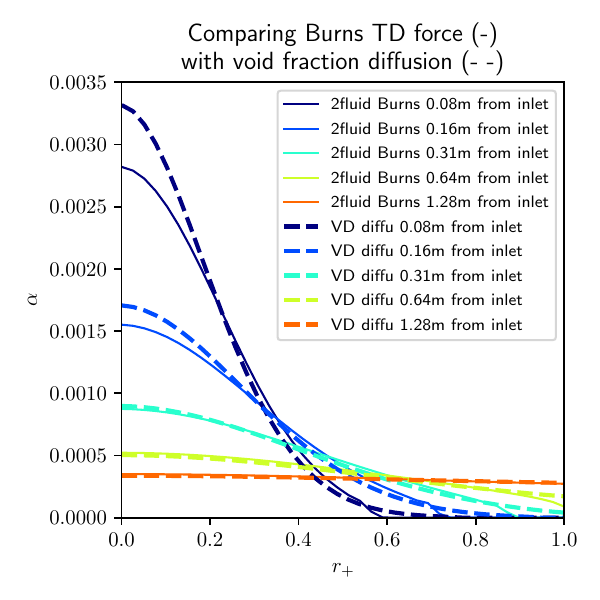}		
	\caption{Comparison between the simulation results obtained with the Burns turbulent dispersion in a 2-fluid model and a void-drift model enforcing passive scalar-like turbulent diffusion.}
	\end{center}
\end{figure}

The results of this simulation are shown in figure~\ref{fig_burns_diffu}. 
There are slight differences between both simulations. 
This is expected, as the conditions in which our approximation to obtain equation~\ref{eq_diffusion_cylinder} is valid aren't completely satisfied. 
In particular, the axial velocities can no longer be considered constant along the width of the pipe. 
Furthermore, we have kept the added mass force, that will play a small part in the center of the pipe as bubbles accelerate after injection.

However, considering the differences in equations solved between the two configurations and the small differences in the outputs, this simulation reinforces the case for the Burns turbulent dispersion to amount to passive scalar turbulent diffusion in fast-flowing pipe flow.

\paragraph{Measures of fluctuations in turbulent flows}

In bubble column experiments where the bubble and liquid velocity fluctuations were both measured~\cite{IMFT2012}, we have $ u_l' < u_g' < 3u_l'$.

If the time scale of velocity fluctuations of the liquid and the gas phase are similar (yet another hypothesis), then $ \nu_l \lesssim \nu_g \lesssim 3\nu_l$. 
As the Burns turbulent dispersion with a Prandtl between .33 and 1 in a flow where bubble distribution is piloted by a drag-dispersion equilibrium would also yield $ \nu_l \lesssim \nu_g \lesssim 3\nu_l$, it would be a relatively good approximation.

\section{Conclusion}

It has been shown that in a system where the turbulent dispersion force is countered by the radial drag force, like a fast-flowing vertical PWR pipe flow, the Burns turbulent dispersion force amounts to passive scalar turbulent diffusion in the radial direction.
This is a useful key to interpret void fraction dispersion results in two-fluid simulations.
Considering the Stokes number of PWR-condition flows, this behavior is in line with what is expected from the flow at low void fractions.

\section{Acknowledgments}

The author thanks Antoine Gerschenfeld, Catherine Colin, Guillaume Bois and Alan Burlot for productive discussions and comments on the draft.

\section{Copyright}

For the purpose of Open Access, a CC-BY public copyright license has been applied by the author to the present document.
\\~\\
\includegraphics[width=3cm]{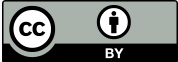}

\bibliographystyle{plain} 
\bibliography{Bibliographie_turb_disp_hal.bib} 

\end{document}